\documentclass[reqno,11pt]{amsart}
\usepackage{graphicx}
\usepackage{slashed}
\usepackage{amscd}
\usepackage{tensor}
\usepackage{amssymb}
\usepackage{esint}
\usepackage{enumitem}
\usepackage{accents}
\usepackage[usenames,dvipsnames]{pstricks}
\usepackage{pst-grad} 
\usepackage{pst-plot} 
\usepackage[mathscr]{eucal}
\numberwithin{equation}{section}
\allowdisplaybreaks[1]

\title[Positive Functionals for Causal Variational Principles]
{Positive Functionals Induced by Minimizers \\ of Causal Variational Principles}

\author[F.\ Finster]{Felix Finster \\ \\ August 2017}
\address{Fakult\"at f\"ur Mathematik \\ Universit\"at Regensburg \\ D-93040 Regensburg \\ Germany}
\email{finster@ur.de}

\newtheorem{Def}{Definition}[section]
\newtheorem{Thm}[Def]{Theorem}
\newtheorem{Prp}[Def]{Proposition}
\newtheorem{Lemma}[Def]{Lemma}
\newtheorem{Remark}[Def]{Remark}

\newcommand{\beq}{\begin{equation}}
\newcommand{\eeq}{\end{equation}}
\newcommand{\Proof}{\begin{proof}}
\newcommand{\QED}{\end{proof} \noindent}
\newcommand{\QEDrem}{\ \hfill $\Diamond$}

\newcommand{\la}{\langle}
\newcommand{\ra}{\rangle}
\newcommand{\llb}{\big\langle\!\langle}
\newcommand{\rrb}{\rangle\!\big\rangle}
\newcommand{\lla}{\langle}
\newcommand{\rra}{\rangle}

\newcommand{\R}{\mathbb{R}}
\newcommand{\1}{\mbox{\rm 1 \hspace{-1.05 em} 1}}

\newcommand{\N}{\mathbb{N}}

\renewcommand{\H}{\mathscr{H}}
\newcommand{\J}{\mathfrak{J}}

\newcommand{\bep}{\begin{pmatrix}}
\newcommand{\enp}{\end{pmatrix}}

\newcommand{\F}{{\mathscr{F}}}

\renewcommand{\L}{{\mathcal{L}}}
\newcommand{\Sact}{{\mathcal{S}}}

\newcommand{\as}{{\mathfrak{a}}}
\newcommand{\bs}{{\mathfrak{b}}}

\newcommand{\Thanks}{
\noindent \thanks}

\DeclareFontFamily{OT1}{rsfso}{}
\DeclareFontShape{OT1}{rsfso}{m}{n}{ <-7> rsfso5 <7-10> rsfso7 <10-> rsfso10}{}
\DeclareMathAlphabet{\myscr}{OT1}{rsfso}{m}{n}

\newcommand\reallywidehat[1]{\arraycolsep=0pt\relax%
\begin{array}{c}
\stretchto{
  \scaleto{
    \scalerel*[\widthof{\ensuremath{#1}}]{\kern-.5pt\bigwedge\kern-.5pt}
    {\rule[-\textheight/2]{1ex}{\textheight}} 
  }{\textheight} %
}{0.5ex}\\           
#1\\                 
\rule{-1ex}{0ex}
\end{array}
}

\DeclareMathOperator{\supp}{supp}

\renewcommand{\u}{\mathfrak{u}}
\renewcommand{\v}{\mathfrak{v}}

\begin{document}
\maketitle

\vspace*{-0.4cm}

\centerline{\em{\rm{\small{\em{Dedicated to the memory of Eberhard Zeidler}}}}}

\begin{abstract}
Considering second variations about a given minimizer of a causal variational principle,
we derive positive functionals in space-time.
It is shown that the strict positivity of these functionals 
ensures that the minimizer is nonlinearly stable
within the class of compactly supported variations with local fragmentation.
As applications, we endow the space of jets in space-time with Hilbert space structures
and derive a positive surface layer integral on solutions of the linearized field equations.
\end{abstract}

\tableofcontents

\section{Introduction} \label{secintro}
Given a minimizer of a variational principle, second variations 
are always non-negative.
This basic observation goes back to Legendre and Jacobi, who used it to analyze the
question whether classical trajectories minimize the action and
geodesics minimize arc length~\cite{goldstine}.
In classical field theory (like electrodynamics or general relativity),
second variations are less useful because in these theories the action is unbounded
from below, so that instead of minimizing one merely seeks for critical points of the action.
However, in the recent theory of causal fermion systems
(see the textbook~\cite{cfs} or the physical introduction~\cite{dice2014}), the action
is indeed bounded from below, and physical space-time should be described by a minimizer.
Therefore, second variations should give rise to positive functionals in space-time.
In special situations and for specific variations (so-called scalar variations;
see Remark~\ref{remscalar} below), such functionals
have already been obtained in~\cite[Lemma~3.5]{support} and~\cite[Sections~4.4 and~4.5]{lagrange}.
Here we give a more general construction which gives rise to two positive functionals.
We show that these functionals describe the local behavior of the action completely, in the sense that
if the functionals are strictly positive, then the minimizer is nonlinearly stable within a 
well-defined general class of variations.
Moreover, it is shown that our positive functionals can be used to endow the
space of jets in space-time with Hilbert space structures. Finally, our functionals give rise to
a positive surface layer integral on solutions of the linearized field equations.

In general terms, in a {\em{causal variational principle}} one minimizes an {\em{action}}~$\Sact$ of the form
\beq \label{Sact} 
\Sact (\rho) = \int_\F d\rho(x) \int_\F d\rho(y)\: \L(x,y) 
\eeq
under variations of the measure~$\rho$, keeping the total volume~$\rho(\F)$ fixed
({\em{volume constraint}}).
In order to keep the presentation self-contained and reasonably simple,
we here restrict attention to the {\em{smooth setting}} as considered in~\cite[Section~3]{jet}
(but all our arguments and results can be generalized
in a straightforward manner to the lower-semicontinuous setting by distinguishing
vector fields pointing in directions in which directional derivatives exist;
for details see~\cite[Section~4]{jet}).
Thus we let~$\F$ be a (possibly {\em{non-compact}}) smooth manifold of dimension~$m \in \N$.
Moreover, the {\em{Lagrangian}}~$\L \in C^\infty(\F \times \F, \R^+_0)$ is given as a smooth non-negative function
which is symmetric (i.e.\ $\L(x,y) = \L(y,x)$ for all~$x,y \in \F$).
Let~$\rho$ be a minimizing measure (for details see Section~\ref{secprelim} below).
We refer to the support of the measure as
\[ \text{space-time} \qquad M:= \supp \rho \:. \]
The notion {\em{causal}} in ``causal variational principles'' refers to the fact that
the Lagrangian induces on~$M$ a causal structure. Namely, two space-time points~$x,y \in M$
are said to be timelike and space-like separated if~$\L(x,y)>0$ and~$\L(x,y)=0$, respectively.
For more details on this notion of causality, its connection to the causal structure
in Minkowski space and to general relativity we refer to~\cite[Chapter~1]{cfs}, \cite{nrstg}
and~\cite[Sections~4.9 and~5.4]{cfs}.

First variations can be computed with the help of the formula
\beq \label{deltaS}
\delta S = 2 \int_\F \ell(x) \:d\delta \rho(x)
\qquad \text{with} \qquad
\ell(x) := \int_\F \L(x,y)\: d\rho(y) - \frac{\nu}{2} \:,
\eeq
where~$\nu$ is an arbitrary real parameter.
The condition that this variation be non-negative for any variation~$\delta \rho$
which respects the volume constraint implies that the function~$\ell$
is minimal and constant on~$M$ (for details see again Section~\ref{secprelim} below).
We always choose the real parameter~$\nu$ such that this constant is zero, i.e.\
\begin{align}\label{ELstrongintro}
\ell|_M \equiv \inf_\F \ell = 0
\end{align}
This means in words that when minimizing the action, the measure~$\rho$ is driven towards the
minima of the function~$\ell$, which in turn are arranged to all have the same value of~$\ell$.
As a result, space-time~$M$ will typically be a discrete or low-dimensional subset of~$\F$
(this picture has been confirmed by numerical and analytic results in~\cite{support}).

For the variations we consider families of measures~$(\tilde{\rho}_\tau)_{\tau \in [0,\tau_{\max})}$
with~$\tau_{\max}>0$ such that~$\tilde{\rho}_0$ coincides with our minimizer~$\rho$.
We now explain how to choose the class of variations.
First of all, the family should satisfy the volume constraint (as will be made precise again in
Section~\ref{secprelim} below). Moreover, if the first variation is strictly positive,
then the second variation need not be considered.
Thus in view of~\eqref{deltaS} and~\eqref{ELstrongintro}, it suffices to consider the
situation that the support of the varied measures lies in the set where~$\ell$
is small, i.e.\ given any~$\varepsilon>0$ we may assume that
\beq \label{suppeps}
\supp \tilde{\rho}_\tau \subset \ell^{-1} \big( [0, \varepsilon) \big) \qquad \text{for all~$\tau \in [0,\tau_{\max})$}\:.
\eeq
There is the subtle issue that the set~$\ell^{-1}([0, \varepsilon))$ may contain points which
are not in a small neighborhood of~$M$, no matter how small we choose~$\varepsilon$.
For example, there may be points~$y \in \F \setminus M$ with~$\ell(y)=0$.
This situation has been analyzed in~\cite[Section~3.5]{lagrange}, and we will
discuss it in Remark~\ref{remaddpoint} below.
Moreover, there might be sequences~$(y_n)_{n \in \N}$ with~$\ell(y_n) \searrow 0$
which have no accumulation points (note that~$\F$ may be non-compact).
This would mean intuitively that ``$\ell$ has a minimum at infinity,'' implying that~\eqref{suppeps} would not
give us control of the support of~$\tilde{\rho}_\tau$ for small~$\tau$.
This case, which seems difficult and somewhat artificial, will not be covered by our analysis.
Instead, we shall only consider variations supported
in a small neighborhood of~$M$, i.e.\ for any open neighborhood~$U$ of~$M$ we assume that
by decreasing~$\tau_{\max}$ we can arrange that
\beq \label{rhocond}
\supp \tilde{\rho}_\tau \subset U \qquad \text{for all~$\tau \in [0,\tau_{\max})$}\:.
\eeq

\begin{figure}%
%
\psscalebox{1.0 1.0} 
{
\begin{pspicture}(-2.1,-1.332472)(13.569116,1.332472)
\definecolor{colour0}{rgb}{0.8,0.8,0.8}
\pspolygon[linecolor=colour0, linewidth=0.02, fillstyle=solid,fillcolor=colour0](5.16583,1.188737)(5.3080525,-1.2779297)(9.663608,-1.3223741)(9.930275,1.3220704)
\pspolygon[linecolor=colour0, linewidth=0.02, fillstyle=solid,fillcolor=colour0](0.010274713,1.3220704)(0.072496936,-1.0334853)(4.321386,-0.9579297)(4.2724967,1.1620703)
\psbezier[linecolor=black, linewidth=0.02, arrowsize=0.05291667cm 2.0,arrowlength=1.4,arrowinset=0.0]{->}(2.9687328,-0.11792968)(2.9891067,-0.021550179)(3.0252702,0.15729474)(3.0147192,0.23887472869060333)(3.004168,0.32045472)(2.9437313,0.4651642)(2.8813858,0.5620703)
\rput[bl](3.1791637,0.07095921){$F_\tau$}
\rput[bl](6.801386,-0.76904076){$F_{2,\tau}$}
\rput[bl](6.8102746,0.124292545){$F_{1,\tau}$}
\psbezier[linecolor=black, linewidth=0.02, arrowsize=0.05291667cm 2.0,arrowlength=1.4,arrowinset=0.0]{->}(6.6531773,0.01095921)(6.673551,0.10733871)(6.709715,0.28618363)(6.6991634,0.3677636175794927)(6.6886125,0.4493436)(6.6281757,0.5940531)(6.56583,0.6909592)
\psbezier[linecolor=black, linewidth=0.02, arrowsize=0.05291667cm 2.0,arrowlength=1.4,arrowinset=0.0]{->}(6.66515,-0.34904078)(6.653665,-0.39208695)(6.657501,-0.49982077)(6.6547194,-0.5147340867363914)(6.651937,-0.5296474)(6.690151,-0.65657914)(6.721386,-0.7757075)
\psbezier[linecolor=white, linewidth=0.04](9.835831,-0.23348524)(10.383583,-0.5184006)(11.0908,-0.42034507)(11.524719,-0.3434852345784503)(11.958638,-0.2666254)(12.590771,-0.015595011)(13.566941,0.090959206)
\rput[bl](3.8680525,0.808737){$\F$}
\rput[bl](5.312497,0.8220703){$\F$}
\pscircle[linecolor=black, linewidth=0.04, fillstyle=solid,fillcolor=black, dimen=outer](2.9213858,-0.23792967){0.08}
\rput[bl](3.0458302,-0.431263){$x \in M$}
\rput[bl](0.09471916,0.38651475){$F_\tau(x) \in \supp \tilde{\rho}_\tau$}
\rput[bl](5.7969413,-0.14681856){$F_{3,\tau}$}
\psbezier[linecolor=black, linewidth=0.02, arrowsize=0.05291667cm 2.0,arrowlength=1.4,arrowinset=0.0]{->}(6.5258303,-0.23654374)(6.4332304,-0.21549748)(6.329547,-0.22425202)(6.2511663,-0.25126301235623033)(6.1727858,-0.278274)(6.085254,-0.3174363)(5.903608,-0.43348524)
\rput[bl](6.6369414,0.728737){$(F_{1,\tau})_* (f_{1,\tau}\, c_1\, \rho)$}
\rput[bl](6.819164,-0.30681857){$x \in M$}
\pscircle[linecolor=black, linewidth=0.04, dimen=outer](2.8191636,0.72207034){0.08}
\pscircle[linecolor=black, linewidth=0.04, fillstyle=solid,fillcolor=black, dimen=outer](6.672497,-0.19348523){0.08}
\pscircle[linecolor=black, linewidth=0.04, dimen=outer](6.4636083,0.8109592){0.08}
\pscircle[linecolor=black, linewidth=0.04, dimen=outer](5.8058305,-0.49570745){0.08}
\pscircle[linecolor=black, linewidth=0.04, dimen=outer](6.779164,-0.9134852){0.08}
\end{pspicture}
}
\caption{Varying the measure with fragmentation.}%
\label{figfragment}%
\end{figure}
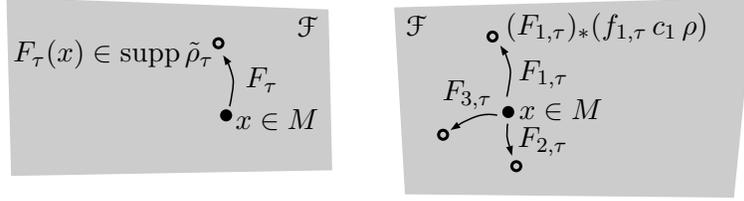

A method for varying measures which has been proven fruitful in~\cite{jet} is to
multiply~$\rho$ by a positive smooth weight function~$f_\tau$ and then to take the push-forward
under a smooth mapping~$F_\tau : M \rightarrow \F$. This leads to the ansatz
(for details see Section~\ref{secJ} below)
\beq \label{Ffvary}
\tilde{\rho}_\tau = (F_\tau)_* \big( f_\tau \,\rho \big) \:.
\eeq
Since multiplying by a positive function leaves the support unchanged,
the support of the measure is transformed only by~$F_\tau$; more precisely,
\[ \supp \tilde{\rho}_\tau = \overline{F_\tau \big( \supp \rho \big)} \:. \]
In particular, for this family of measures, the support changes smoothly in~$\tau$.
For example, if~$M$ is discrete, then the measures~$\rho_\tau$ will necessarily also
be discrete (see the left of Figure~\ref{figfragment} for the example of a Dirac measure).
However, this is a much stronger condition than~\eqref{rhocond}, according to which
a measure supported at a point~$x \in M$
could ``disintegrate'' in the variation and become for example a weighted sum of Dirac measures supported
in a small neighborhood of~$x$ (see the right of Figure~\ref{figfragment}).
In order to take such effects into account, 
using the notion first introduced in~\cite[Section~5]{perturb}, we want to allow for the possibility
of a {\em{fragmentation}} of the measure~$\rho$.
In mathematical terms, fragmentation is described as follows: Given a parameter~$L \in \N$,
we choose weight functions~$c_\as : M \rightarrow \R^+$ for~$a=1,\ldots, L$
which add up to one (for details see Section~\ref{secfragment})
\[ \sum_{\as=1}^L c_\as(x) = 1 \quad \text{for all~$x \in M$}\:. \]
Then we vary each of the measures~$c_\as \,\rho$ similar to~\eqref{Ffvary}
by functions~$(f_{\as,\tau}, F_{\as,\tau})$ (see again the right of Figure~\ref{figfragment}),
\beq \label{rhotildea}
\tilde{\rho}_\tau = \sum_{\as=1}^L (F_{\as,\tau})_* \big(f_{\as,\tau} \,c_\as \,\,\rho \big) \:.
\eeq
Since~$L$ can be chosen arbitrarily large, the resulting variations with fragmentation
can be used to approximate all smooth variations which satisfy~\eqref{rhocond}.

For technical simplicity, we restrict attention to variations which are 
{\em{compactly supported}} (i.e.\ trivial outside a compact set; for details see Section~\ref{secJ}).
Then the variation of the measure~\eqref{Ffvary} is described infinitesimally by a
pair of a real-valued function and a vector field on~$M$ (for details see Section~\ref{secl}),
both with compact support,
\[ (a,u) := \frac{d}{d\tau} \big( f_\tau, F_\tau \big)\big|_{\tau=0} \;\in\;
\J^\infty_0 := C^\infty_0(M, \R) \oplus C^\infty_0(M, T\F) \:. \]
The pair~$\u := (a,u)$ is referred to as a {\em{jet}}.
We denote the linear combination of multiplication and directional derivative by~$\nabla$, i.e.
\beq \label{Djet}
\nabla_{\u} g(x) := a(x)\, g(x) + \big(D_u g \big)(x)
\eeq
(where~$g$ is a smooth function on~$\F$).
Then the EL equations~\eqref{ELstrongintro} evaluated in a linear Taylor approximation
on~$M$ give rise to the so-called {\em{weak EL equations}}
\beq \label{ELweak2}
\nabla_\u \ell|_M = 0 \qquad \text{for all~$\u \in \J^\infty_0$}\:.
\eeq

We now state our main result:
\begin{Thm} \label{thmstable}
Let~$\rho$ be a minimizer of a causal variational principle in the
non-compact smooth setting (for details see Section~\ref{secprelim}).
Then the following two quadratic functionals on~$\J^\infty_0$ are positive:
\begin{gather}
\int_M \nabla^2 \ell|_x(\u,\u)\: d\rho(x) \geq 0 \label{Fpos1} \\
\int_M d\rho(x) \int_M d\rho(y) \:\nabla_{1,\u} \nabla_{2,\u} \L(x,y) 
+ \int_M \nabla^2 \ell|_x(\u,\u)\: d\rho(x) \geq 0 \:. \label{Fpos2}
\end{gather}

Conversely, assume that~$\rho$ is a Radon measure which satisfies~\eqref{ELweak2}
as well as the inequalities~\eqref{Fpos1} and~\eqref{Fpos2}.
If the inequality~\eqref{Fpos2} is strict for every non-zero~$\u \in \J^\infty_0$, then~$\rho$
is an isolated local minimum within the class of compactly supported variations with local
fragmentation~\eqref{rhotildea}.
\end{Thm}

The paper is organized as follows. In Section~\ref{secprelim} we recall
the necessary preliminaries on causal variational principles in the smooth setting.
In Section~\ref{secl}, a positive functional is derived from the Hessian of~$\ell$.
Section~\ref{secJ} is devoted to another positive functional obtained by
second variations generated by jets.
In Section~\ref{secfragment} the stability analysis is extended such as to
allow for a fragmentation of the measure~$\rho$.
This will conclude the proof of Theorem~\ref{thmstable}.
The last two sections are devoted to the applications:
In Section~\ref{sechilbert} it is shown that the positive functionals 
endow the jet spaces with Hilbert space structures. Finally,
in Section~\ref{secosi} a positive surface layer integral is derived.

I want to close with a personal remark. This paper is dedicated to the memory of Eberhard Zeidler,
whom I admire both for his astounding knowledge and his deeply honest and
amiable personality.
I am very grateful and thankful for his guidance and advice.
When I was post-doc at the Max Planck Institute for Mathematics in the Sciences
in Leipzig from 1998 to 2002, we had long conversations in which he helped me to
find the right direction of my research. In particular, I vividly remember that
he encouraged me in my study of the interacting Dirac sea 
by telling me not to aim for short-term success or recognition by
a scientific community. He told me that in the long run, thinking thoroughly
on a problem would pay off. Taking this advice seriously led me to the
physical theory of causal fermion systems and to the mathematical
setting of causal variational principles, a certain aspect of which is presented here.

\section{Preliminaries: Minimizers of Causal Variational Principles} \label{secprelim}
Let~$\F$ be a (possibly {\em{non-compact}}) smooth manifold of dimension~$m \in \N$.
Let~$\L \in C^\infty(\F \times \F, \R^+_0)$ be a non-negative smooth function which is symmetric, i.e.\
\[ \L(x,y) = \L(y,x) \qquad \text{for all~$x,y \in \F$}\:. \]
We let~$\rho$ be a {\em{Radon measure}} on~$\F$ (i.e.\ a regular Borel measure
with~$\rho(K)<\infty$ for any compact~$K \subset \F$, where
by a measure we always mean a {\em{positive}} measure; for preliminaries see for
example~\cite{halmosmt} or~\cite{bogachev}).
Moreover, we assume that~$\rho$ satisfies the following technical assumption:
\begin{itemize}[leftmargin=2em]
\item[(a)] For all~$x \in \F$, the functions~$\L(x,.)$ and~$\partial_{x^j} \L(x,.)$ are $\rho$-integrable, giving
smooth and bounded functions on~$\F$. Moreover, partial derivatives and $y$-integration may be interchanged, i.e.\
\beq \label{commute}
\frac{\partial}{\partial x^j} \int_M \L(x,y)\: d\rho(y) = \int_M \frac{\partial \L(x,y)}{\partial x^j}\: d\rho(y)\:.
\eeq
\label{Cond4}
\end{itemize}
If the total volume~$\rho(\F)$ is finite, the {\em{causal variational principle}} is to minimize the
action~\eqref{Sact} under variations of the measure~$\rho$ (which
do not need to satisfy~(a)), keeping the total volume~$\rho(\F)$ fixed
({\em{volume constraint}}).
If~$\rho(\F)$ is infinite, however, it is not obvious how to implement the volume constraint,
making it necessary to proceed as follows:
Let~$\tilde{\rho}$ be another Borel measure on~$\F$
(which again does not need not satisfy~(a)) which has the properties
\beq \label{totvol}
\big| \tilde{\rho} - \rho \big|(\F) < \infty \qquad \text{and} \qquad
\big( \tilde{\rho} - \rho \big) (\F) = 0
\eeq
(where~$|.|$ denotes the total variation of a measure;
see~\cite[\S28]{halmosmt} or~\cite[Section~6.1]{rudin}).
Then the difference of the actions as given by
\beq \label{integrals}
\begin{split}
\big( &\Sact(\tilde{\rho}) - \Sact(\rho) \big) = \int_\F d(\tilde{\rho} - \rho)(x) \int_\F d\rho(y)\: \L(x,y) \\
&\quad + \int_\F d\rho(x) \int_\F d(\tilde{\rho} - \rho)(y)\: \L(x,y) 
+ \int_\F d(\tilde{\rho} - \rho)(x) \int_\F d(\tilde{\rho} - \rho)(y)\: \L(x,y)
\end{split}
\eeq
is well-defined (for details see~\cite[Lemma~2.1]{jet}). 
The measure~$\rho$ is said to be a {\em{minimizer}} of the causal action
if the difference~\eqref{integrals} is non-negative for all~$\tilde{\rho}$ satisfying~\eqref{totvol},
\[ \big( \Sact(\tilde{\rho}) - \Sact(\rho) \big) \geq 0 \:. \]

We now state the Euler-Lagrange (EL) equations as derived
in~\cite[Lemma~2.3]{jet} (by adapting~\cite[Lemma~3.4]{support} to the non-compact setting).
\begin{Lemma} (The Euler-Lagrange equations) \label{lemmaEL}
Let~$\rho$ be a minimizer of the causal action. Then
for a suitable value of the real parameter~$\nu$, the smooth function~$\ell$ defined by
\beq \label{elldef}
\ell(x) = \int_\F \L(x,y)\: d\rho(y) - \frac{\nu}{2} \::\: \F \rightarrow \R
\eeq
satisfies the equation
\begin{align}\label{ELstrong}
\ell|_{\supp \rho} \equiv \inf_\F \ell = 0 \: .
\end{align}
\end{Lemma} \noindent
We remark that~$\nu$ can be understood as the Lagrange multiplier describing the volume constraint;
see~\cite[\S1.4.1]{cfs}.

\section{Positivity of the Hessian of~$\ell$} \label{secl}
Let~$\rho$ be a minimizer of the causal action. 
According to the EL equations~\eqref{ELstrong}, the function~$\ell$ is minimal on~$M$.
This clearly implies that its Hessian (as computed in any chart) is positive semi-definite, i.e.
\beq \label{hessian}
D^2 \ell(x) \geq 0 \qquad \text{for all~$x \in M := \supp \rho$} \:.
\eeq
This is the first non-negative quantity obtained from the fact that~$\rho$ is a minimizer.

In order to clarify the connection to other non-negative functionals below,
it is preferable to rewrite~\eqref{hessian} in the {\em{jet formalism}}, which we now recall
in the smooth setting.
We denote the smooth vector fields on~$\F$
restricted to~$M$ by~$C^\infty(M, T\F)$ (where smoothness is defined by the
condition that every~$u \in C^\infty(M, T\F)$ can be extended to a smooth vector field on~$\F$).
We introduce the smooth {\em{jet space}} $\J^\infty$ by
\[ \J^\infty = C^\infty(M, \R) \oplus C^\infty(M, T\F) \:. \]
The compactly supported jets are denoted by~$\J^\infty_0$.
For a jet~$\u = (a, u) \in \J^\infty$ with {\em{scalar component}}~$a$ and {\em{vector component}}~$u$,
we define~$\nabla_\u$ as the linear combination of
scalar multiplication and directional derivative~\eqref{Djet}.
Then the EL equations~\eqref{ELstrong} clearly
imply the so-called {\em{weak EL equations}}~\eqref{ELweak2}.
The main difference to the strong EL equations is that in the weak formulation
we restrict attention to the behavior of~$\ell$ in an infinitesimal neighborhood
of space-time~$M$ (but in contrast to weak solutions of PDEs, ``weak'' does not refer
to multiplying by a test function and integrating).
Working with jets has the advantage that the two
equations~$\ell|_M=0$ and~$D\ell|_M=0$ are combined in a single equation.

Adapting the notation~\eqref{Djet} to second derivatives, we set
\[ \nabla^2 \ell|_x(\u,\u) := a(x)^2\, \ell(x) + 2 \,a(x)\, D_u \ell(x) + D^2 \ell|x(u,u) \:. \]
In view of the weak EL equations~\eqref{ELweak2}, the zero and first order derivatives 
in this equation vanish for all~$x \in M$. Therefore, inequality~\eqref{hessian} implies that
\[ \nabla^2 \ell|_x(\u,\u) \geq 0 \qquad \text{for all~$x \in M$}\:. \]
Integrating over~$M$ gives the following result:
\begin{Prp} \label{prppos1} Let~$\rho$ be a minimizer of the causal action. Then
\[ \int_M \nabla^2 \ell|_x(\u,\u)\: d\rho(x) \geq 0 \qquad \text{for all~$\u \in \J^\infty_0$} \:. \]
\end{Prp} \noindent

\section{Positivity of Second Variations Generated by Jets} \label{secJ}
In this section we analyze second variations for a 
a special class of variations of the measure~$\rho$ to obtain another positive
functional on jets.
Similar as in~\cite[Section~3]{jet} we consider measures of the form
\beq \label{paramtilrho}
\tilde{\rho}_\tau = (F_\tau)_* \big( f_\tau \,\rho \big) \qquad \text{for $\tau \in (-\tau_{\max}, \tau_{\max})$}
\eeq
with smooth mappings
\[ f \in C^\infty((-\tau_{\max}, \tau_{\max}) \times M, \R^+) \qquad \text{and} \qquad
F \in C^\infty((-\tau_{\max}, \tau_{\max}) \times M, \F) \:, \]
where the star denotes the push-forward measure defined
by~$((F_\tau)_*\mu)(\Omega) = \mu ( F_\tau^{-1} (\Omega))$
(where~$\Omega \subset \F$; for basics see for example~\cite[Section~3.6]{bogachev}).
We assume that for~$\tau=0$ the variation is trivial,
\beq \label{triv0}
f_0 \equiv 1 \qquad \text{and} \qquad F_0 \equiv \1\:.
\eeq
Moreover, we assume that~$F_\tau$ and~$f_\tau$ are {\em{compactly supported}},
meaning that they are trivial outside a compact set~$K \subset M$, i.e.
\beq \label{triv}
f_\tau|_{M \setminus K} \equiv 1 \qquad \text{and} \qquad F_\tau|_{M \setminus K} \equiv \1\:.
\eeq
Finally, in order to satisfy the volume constraint on the right side of~\eqref{totvol}, we assume that
\beq \label{vol0}
\int_K f_\tau(x)\: d\rho(x) = \rho(K) \qquad \text{for all~$\tau \in (-\tau_{\max}, \tau_{\max})$}\:.
\eeq
Then the transformation~\eqref{paramtilrho} is described infinitesimally by a smooth
and compactly supported jet,
\[ \u =(a,u) := \big( \dot{f}_0, \dot{F}_0 \big) \in \J^\infty_0 \:. \]
Moreover, differentiating the volume constraint~\eqref{vol0} gives
\beq \label{vol1}
\int_K a(x)\: d\rho(x) = 0\:.
\eeq

We now compute the first and second variation of the action. Combining~\eqref{integrals}
with the definition of the push-forward measure, we obtain
\begin{align*}
\Sact\big(\tilde{\rho}_\tau \big) - \Sact(\rho) &= 2 \int_K d\rho(x) \int_{M \setminus K} d\rho(y)\: 
\Big( f_\tau(x)\: \L\big(F_\tau(x), y \big) - \L(x,y) \Big) \\
&\qquad + \int_K d\rho(x) \int_K d\rho(y)\:\Big( f_\tau(x)\: f_\tau(y) \:\L\big(F_\tau(x), F_\tau(y) \big) - \L(x,y) \Big) \:.
\end{align*}
Then the first variation vanishes,
\[ \frac{d}{d\tau} \Sact\big(\tilde{\rho}_\tau \big) \Big|_{\tau=0} =
2 \int_K d\rho(x) \int_M d\rho(y) \:\nabla_{1,\u} \L(x,y)
= 2 \int_K \nabla_\u \Big( \ell(x) + \frac{\nu}{2} \Big)\: d\rho(x) = 0 \:, \]
where in the last step we used~\eqref{ELweak2} and~\eqref{vol1}
(and~$\nabla_1$ denotes the partial derivative acting on the first argument of the Lagrangian).
Moreover, the second variation is
computed by
\begin{align*}
&\frac{d^2}{d\tau^2} \Sact\big(\tilde{\rho}_\tau \big) \Big|_{\tau=0} =
2 \int_K d\rho(x) \int_K d\rho(y) \:\nabla_{1,\u} \nabla_{2,\u} \L(x,y) \\
&+ 2 \int_K d\rho(x) \int_M d\rho(y) \:\Big( a(x)\: D_{1,u} \L(x,y) + D_{1,u} D_{1,u} \L(x,y) + 
\big( \ddot{f}_0(x) + D_{1,\ddot{F}_0} \big) \L(x,y) \Big) \:.
\end{align*}
In the last line we can carry out the $y$-integration using~\eqref{elldef}. Again
combining the EL equations~\eqref{ELweak2} with~\eqref{vol1}
(and a similar formula for~$\ddot{f}$), we obtain the simple formula
\beq
\frac{1}{2}\:\frac{d^2}{d\tau^2} \Sact\big(\tilde{\rho}_\tau \big) \Big|_{\tau=0} =
\int_K d\rho(x) \int_K d\rho(y) \:\nabla_{1,\u} \nabla_{2,\u} \L(x,y) 
+ \int_K \nabla^2 \ell|_x(\u,\u)\: d\rho(x)\:. \label{var2}
\eeq
Since~$\rho$ is a minimizer and the first variation vanishes,
the second variation is necessarily non-negative, giving rise
to the inequality
\beq \label{varint}
\int_M d\rho(x) \int_M d\rho(y) \:\nabla_{1,\u} \nabla_{2,\u} \L(x,y) 
+ \int_M \nabla^2 \ell|_x(\u,\u)\: d\rho(x) \geq 0\:,
\eeq
subject to the condition that the jet~$\u$ must satisfy the volume constraint~\eqref{vol1}.
In the next proposition we remove this condition with a limiting procedure:
\begin{Prp} \label{prppos2} Let~$\rho$ be a minimizer of the causal action. Then
\[ \int_M d\rho(x) \int_M d\rho(y) \:\nabla_{1,\u} \nabla_{2,\u} \L(x,y) 
+ \int_M \nabla^2 \ell|_x(\u,\u)\: d\rho(x) \geq 0 \qquad \text{for all~$\u \in \J^\infty_0$}\:. \]
\end{Prp}
\Proof Let~$\u=(a,u) \in \J^\infty_0$ be a jet which violates
the volume constraint~\eqref{vol1}. Then, choosing a compact set~$\Omega \subset M$
with~$\rho(\Omega)>0$, the jet~$\hat{\u} := (\hat{a}, u)$ with
\beq \label{cdef}
\hat{a}(x) = a(x) - c(\Omega) \,\chi_\Omega(x) \qquad \text{and} \qquad
c(\Omega) := \frac{1}{\rho(\Omega)} \:\int_\Omega a(x)\: d\rho(x)
\eeq
(where~$\chi_\Omega$ is the characteristic function) satisfies the volume constraint.
Choosing the scalar variation~$f_\tau = (1-\tau) + \tau \hat{a}$ and
a family of diffeomorphisms~$F_\tau$ with~$\dot{F}_0 = u$, we obtain a variation
which satisfies the volume constraint (note that~$\ddot{f}=0$).
Clearly, due to the characteristic function, the jet~$\hat{\u}$ is no longer smooth,
but it has again compact support, and an approximation argument using Lebesgue's dominated convergence
theorem shows that the inequality~\eqref{varint} also holds for~$\hat{\u}$.
Expanding in powers of~$c$, we thus obtain the inequality
\begin{align*}
0 &\leq \int_K d\rho(x) \int_K d\rho(y) \:\nabla_{1,\u} \nabla_{2,\u} \L(x,y) 
+ \int_K \nabla^2 \ell|_x(\u,\u)\: d\rho(x) \\
&\qquad - 2c \int_M d\rho(x) \int_K d\rho(y) \:\chi_\Omega(x) \nabla_{2,\u} \L(x,y) \\
&\qquad + c^2 \int_M d\rho(x) \int_M d\rho(y) \:\chi_\Omega(x) \, \chi_\Omega(y)\: \L(x,y) \\
&\qquad + \int_M \Big( -2 c \,\chi_\Omega(x)\: \nabla_\u \ell(x) + c^2\: \chi_\Omega(x)^2\: \ell(x) \Big)\: d\rho(x)
\end{align*}
(the integrand in the last line arises from the contributions to~$\nabla^2 \ell|x(\u,\u)$
involving the scalar components of the jets).
The last line vanishes due to the weak EL equations~\eqref{ELweak2}. Hence
\begin{align*}
&\int_K d\rho(x) \int_K d\rho(y) \:\nabla_{1,\u} \nabla_{2,\u} \L(x,y) + \int_K \nabla^2 \ell|_x(\u,\u)\: d\rho(x)\\
&\geq 2c \int_K d\rho(x) \int_\Omega d\rho(y) \:\nabla_{1,\u} \L(x,y) 
-c^2 \int_K d\rho(x) \int_K d\rho(y) \: \L(x,y) =: A(\Omega) \:.
\end{align*}

We now let~$(\Omega_n)_{n \in \N}$ be an exhaustion of~$M$ by compact sets.
We distinguish the two cases when~$\rho(M)$ is finite and infinite and treat these cases separately.
If the total volume~$\rho(M)$ is finite, one can take the limit~$n \rightarrow \infty$ with
Lebesgue's dominated convergence theorem to obtain
\begin{align*}
\lim_{n \rightarrow \infty} \int_K &d\rho(x) \int_{\Omega_n} d\rho(y) \:\nabla_{1,\u} \L(x,y) =
\int_K d\rho(x) \int_M d\rho(y) \:\nabla_{1,\u} \L(x,y) \\
&\!\!\overset{\eqref{commute}}{=} \int_K \nabla_{\u} \bigg( \ell(x) + \frac{\nu}{2} \bigg) \:d\rho(x)
= \frac{\nu}{2} \int_K a(x)\: d\rho(x) \\
\lim_{n \rightarrow \infty} A(\Omega_n) &= 2\,c(M) \:\frac{\nu}{2} \int_K a(x)\: d\rho(x) - c(M)^2 \, \rho(M)\: \frac{\nu}{2} \\
&= \frac{\nu}{2 \rho(M)} \:\left( \int_K a(x)\: d\rho(x) \right)^2 \geq 0\:,
\end{align*}
where in the last line we substituted the value of~$c(M)$ in~\eqref{cdef}.

In the remaining case that the volume~$\rho(M)$ is infinite, we estimate the terms as follows,
\begin{align*}
&c(\Omega_n)^2 \int_K d\rho(x) \int_K d\rho(y) \: \L(x,y) \\
&\quad \leq c(\Omega_n)^2 \int_K d\rho(x) \int_M d\rho(y) \: \L(x,y)
= c(\Omega_n)^2 \: \frac{\nu}{2}\: \rho(K) \rightarrow 0 \\
&\int_K d\rho(x) \int_{\Omega_n} d\rho(y) \:\nabla_{1,\u} \L(x,y)
\rightarrow \int_K d\rho(x) \int_M d\rho(y) \:\nabla_{1,\u} \L(x,y) \\
&\quad \!\!\overset{\eqref{commute}}{=} \int_K \nabla_{\u} \bigg( \ell(x) + \frac{\nu}{2} \bigg) \:d\rho(x)
= \frac{\nu}{2} \int_K a(x)\: d\rho(x) \:.
\end{align*}
As a consequence, $A(\Omega_n)$ converges to zero as~$n \rightarrow \infty$.
This concludes the proof.
\QED

We close this section with two remarks.

\begin{Remark} {\bf{(scalar jets)}} \label{remscalar} {\em{
In the special case~$\u=(a,0)$ of a purely {\em{scalar jet}}, the statement of Proposition~\ref{prppos2}
simplifies to the inequality
\beq \label{scalpos}
\int_M d\rho(x) \int_M d\rho(y) \:a(x)\: a(y) \:\L(x,y) \geq 0 \qquad \text{for all~$a \in C^\infty_0(M,\R)$}\:.
\eeq
In the compact setting, this inequality was already derived in~\cite[Lemma~3.5]{support}.
It is a main ingredient in the analysis of the singular support in~\cite[Section~3]{support}
(see~\cite[Theorems~3.16 and~3.18]{support}).
Furthermore, in~\cite[Theorem~3.16]{lagrange} a similar positivity statement was
derived in the setting of causal fermion systems.

Moreover, second variations are studied in~\cite[Section~4.4]{jet},
and a stability result is obtained (see~\cite[Proposition~4.10]{jet}).
In this analysis, only variations~$(\tilde{\rho}_\tau)_{\tau \in (-\tau_{\max}, \tau_{\max})}$ are considered
where~$\mu_\tau := \tilde{\rho}_\tau - \rho$ is a smooth curve in the
Banach space of signed measures of bounded total variation.
In particular, the derivatives~$\dot{\tilde{\rho}}_0$ and~$\ddot{\tilde{\rho}}_0$
are assumed to be signed measures. Such variations do not include
variations of the form~\eqref{paramtilrho}
(because for example the infinitesimal transport of a Dirac measure
gives a distributional derivative of a Dirac measure, which is {\em{not}}
a signed measure).
This is the reason why, using the language of jet spaces, only the scalar component
of the jets comes into play. As a consequence, the stability result follows already
by analyzing the inequality for scalar jets~\eqref{scalpos}.

To summarize, the new feature in Proposition~\ref{prppos2} is the
inclusion of the vector component of the jets. This is
a major improvement which gives much more information, in particular if the dimension of~$\F$ is large.
}} \QEDrem
\end{Remark}

\begin{Remark}  {\bf{(adding points to the support)}} \label{remaddpoint} {\em{
We now discuss the situation that there is a point~$y \in \F \setminus M$ with~$\ell(y)=0$.
In the setting of causal fermion systems, this case was treated in~\cite[Section~3.5]{lagrange}
working with moment measures. For self-consistency, we here repeat the argument in
our setting and explain why the jet formalism does not make it possible to improve the result.

More generally, let~$A \subset \F$ be a compact set with
\[ A \cap M = \varnothing \qquad \text{and} \qquad \ell|_A \equiv 0 \]
(a typical choice is~$A=\{y\}$). We choose a measure~$\mu$ on~$A$ with~$\mu(A)<\infty$.
Moreover, we choose a function~$a \in C^\infty_0(M \cup A, \R^+_0)$ with
\[ a|_A \geq 0 \qquad \text{and} \qquad \int_F a(x)\: d(\rho+\mu)(x) = 0 \:. \]
We now consider the variation
\[ \tilde{\rho}_\tau = (1-\tau\, a)\: \rho + \tau \,a\: \mu \,\qquad \tau \in [0,1) \:. \]
Then the first variation vanishes because~$\ell$ vanishes identically on the support of~$a$.
Hence the second variation must be non-negative. Evaluating this condition gives
the result of~\cite[Theorem~3.17]{lagrange}.

In order to analyze the effect of a vector components of the jets, we consider the more general
variation
\beq \label{rhoaddpoint}
\tilde{\rho}_\tau = (F_\tau)_*\big( (1-\tau\, a)\: \rho + \tau \,a\: \mu \big)
\eeq
with a mapping~$F_\tau \in C^\infty(M \cup A, \F)$ which we again assume to be trivial
outside a compact set. Infinitesimally, this variation is described by the jet
\[ \u = (a,\dot{F}_0) \in C^\infty_0(M \cup A, \R^+_0) \oplus C^\infty_0(M \cup A, T\F) \:. \]
The first variation again vanishes. Thus the second variation must again be non-negative.
On the set~$A$, the second variation necessarily involves the scalar component
(because without the scalar component, $\tilde{\rho}_\tau$ vanishes on~$A$).
As a consequence, the second variation involves the vector component on~$A$ at most linearly.
Therefore, using a block matrix notation where the first component contains the jets in~$M$ as well as the scalar
component in~$A$ and the second component contains the vector component in~$A$,
the second variations are described by a bilinear form with the structure
\[ \la \u, \begin{pmatrix} L_{11} & L_{12} \\ L_{21} & 0 \end{pmatrix} \:\u \ra \]
with suitable operators~$L_{ij}$. The positivity of this bilinear form
is equivalent to the positivity of the upper left matrix entry.
This shows that second variations of the form~\eqref{rhoaddpoint}
do not give more information than~\cite[Theorem~3.17]{lagrange}
and Proposition~\ref{prppos2} above. }} \QEDrem
\end{Remark}

\section{Stability under Variations with Local Fragmentation} \label{secfragment}
We now consider variations of the general form~\eqref{rhotildea},
where the weight functions~$c_\as$ are normalized at every point, i.e.\
\beq \label{constraint}
c_\as(x) \geq 0 \qquad \text{and} \qquad \sum_{\as=1}^L c_\as(x) = 1 \quad \text{for all~$x \in M$}\:.
\eeq
In order to keep the setting as simple as possible, we assume that the functions~$c_\as$
are smooth.
Moreover, similar to~\eqref{triv0} and~\eqref{triv}, we assume that
that for every~$\as$, the transformation is trivial for~$\tau=0$
and is trivial outside a compact set~$K$.
Then our variation is described infinitesimally by jets~$\u_\as = (b_\as, u_\as) \in \J^\infty_0$.
In analogy to~\eqref{vol1}, the volume constraint becomes
\[ \int_K \sum_{\as=1}^L c_\as(x)\: b_\as(x) \:d\rho(x) =  0\:. \]

We remark that we here consider {\em{local}} fragmentations, meaning that
we allow the weights~$c_\as$ to be functions of~$x$.
This is different from the procedure in~\cite[Section~5]{perturb}, where the weights~$c_\as$
were real parameters. However, this is not an important difference, because 
after approximating the weight functions~$c_\as(x)$ by locally constant
functions taking values in~$\N/\tilde{L}$ with large~$\tilde{L}$
and by increasing the number of subsystems to~$\tilde{L}$,
one can approximate a variation involving local fragmentation by
variations with constant weights~$c_\as = 1/\tilde{L}$.
Therefore, considering local fragmentation is mainly a matter of convenience and
has the advantage of giving a somewhat different perspective.

The variation of the action can be computed similar as in Section~\ref{secJ}.
The first variation again vanishes (as is obvious by linearity in the jets).
When computing the second variation, the terms involving up to first derivatives of~$\L$ again 
drop out. We thus obtain in generalization of~\eqref{var2}
\begin{align*}
\frac{1}{2}\:\frac{d^2}{d\tau^2} \Sact\big(\tilde{\rho}_\tau \big) \Big|_{\tau=0} &=
\int_K d\rho(x) \int_K d\rho(y) \sum_{\as, \bs=1}^L c_\as(x)\, c_\bs(y)\:\nabla_{1,\u_\as} \nabla_{2,\u_\bs} \L(x,y) \\
&\qquad + \int_K \sum_{\as=1}^L c_\as(x)\: \nabla^2 \ell|_x(\u_\as,\u_\as)\: d\rho(x) \:.
\end{align*}
In order to clarify the scaling behavior in the weights~$c_\as$, it is useful to
transform the jets according to~$\u_\as \rightarrow \u_\as/c_\as$
(thus we multiply both the scalar and vector components of the jet~$\u$ by the same function~$c_\as(x)$;
this is admissible because this function is smooth and bounded,
so that multiplying by it again gives a jet in~$\J^\infty_0 := C^\infty_0(M, \R) \oplus C^\infty_0(M, T\F)$).
We thus obtain the inequality
\beq \begin{split}
0 \leq \frac{1}{2}\:\frac{d^2}{d\tau^2} \Sact\big(\tilde{\rho}_\tau \big) \Big|_{\tau=0} &=
\int_K d\rho(x) \int_K d\rho(y) \sum_{\as, \bs=1}^L \nabla_{1,\u_\as} \nabla_{2,\u_\bs} \L(x,y)\\
&\qquad + \int_K \sum_{\as=1}^L \frac{1}{c_\as(x)}\: \nabla^2 \ell|_x(\u_\as,\u_\as)\: d\rho(x) \:,
\end{split} \label{var2b}
\eeq
where the weights appear only in the last summand.
Now it is apparent that the positivity of the second variation yields the
positivity statements of both Section~\ref{secl} and Section~\ref{secJ}.
Indeed, in the case~$L=1$ of one subsystem we clearly get back to the setting
of Section~\ref{secJ}. Moreover, in the limiting case~$c_1 \searrow 0$
and~$c_2=\ldots=c_L \approx (L-1)^{-1}$, the factor~$c_1^{-1}$ in~\eqref{var2b}
diverges, implying that
\[ \int_K \nabla^2 \ell|_x(\u_1,\u_1)\: d\rho(x) \geq 0 \:. \]
We thus recover the result of Proposition~\ref{prppos1}.

In order to understand what the inequality~\eqref{var2b} means, it is helpful to
choose the weights~$c_\as$ at every point in such a way that the last summand in~\eqref{var2b}
becomes as small as possible, because then the inequality in~\eqref{var2b} gives
most information. Abbreviating the integrand by~$A_\as = \nabla^2 \ell|_x(\u_\as,\u_\as)\geq 0$, our task is to
\beq \label{minimize}
\text{minimize} \quad \sum_{\as=1}^L \frac{A_\as}{c_\as}
\eeq
under the constraints~\eqref{constraint}. The method of Lagrange multipliers gives
\[ c_\as = \sqrt{\frac{A_\as}{\lambda}} \qquad \text{with} \qquad 
\lambda = \Big( \sum_\as \sqrt{A_\as} \Big)^2 \:. \]
A direct computation shows that this choice of weights satisfies~\eqref{constraint}.
Moreover, the following consideration shows that this choice of weights
indeed realizes the minimum: Clearly, if one of the parameters~$A_\as$ vanishes,
then the minimum is attained in the limiting case~$c_\as \searrow 0$.
With this in mind, we may assume that all the~$A_\as$ are strictly positive.
Next, at the boundary of the admissible
range for the parameters~$\{c_1, \ldots, c_L\}$ one or several of the coefficients~$c_\as$
vanish. As a consequence, the expression in~\eqref{minimize} tends to $+\infty$ at the boundary
of the admissible parameter range.
It follows that the only interior critical point is the absolute minimum.

Substituting the obtained values for~$c_\as$ into~\eqref{var2b}, we obtain the inequality
\beq \begin{split}
0 &\leq \int_K d\rho(x) \int_K d\rho(y) \sum_{\as, \bs=1}^L \nabla_{1,\u_\as} \nabla_{2,\u_\bs} \L(x,y)\\
&\qquad + \int_K \bigg( \sum_{\as=1}^L  \sqrt{ \nabla^2 \ell|_x(\u_\as,\u_\as) } \bigg)^2 \: d\rho(x) \:.
\end{split} \label{var2c}
\eeq
The inequality~\eqref{var2b} holds for all choices of the weight functions if and only if~\eqref{var2c} holds.

From the inequality~\eqref{var2c} one can read off our main stability result:

\Proof[Proof of Theorem~\ref{thmstable}.]
The inequalities~\eqref{Fpos1} and~\eqref{Fpos2} were derived in Propositions~\ref{prppos1}
and~\ref{prppos2}.
In order to derive the stability statement, let~$\rho$ be a Radon measure
which satisfies~\eqref{ELweak2} as well as the inequalities~\eqref{Fpos1} and~\eqref{Fpos2}.
Moreover, assume that the inequality~\eqref{Fpos2} is strict for every non-zero~$\u \in \J^\infty_0$.

Let~$(\tilde{\rho}_\tau)_{\tau \in (-\tau_{\max}, \tau_{\max})}$ be a non-trivial
variation with local fragmentation~\eqref{rhotildea}.
The inequality~\eqref{Fpos1} allows us to introduce a seminorm on the jets by
\beq \label{seminorm}
\| \u(x) \| := \sqrt{ \nabla^2 \ell|_x(\u,\u) }\:.
\eeq
Using the strict inequality in~\eqref{Fpos2}, we obtain
\begin{align*}
-&\int_K d\rho(x) \int_K d\rho(y) \sum_{\as, \bs=1}^L \nabla_{1,\u_\as} \nabla_{2,\u_\bs} \L(x,y) \\
&< \int_K \nabla^2 \ell|_x\bigg( \sum_{\as=1}^L  \u_\as, \sum_{\bs=1}^L  \u_\bs \bigg)
= \int_K \bigg\| \sum_{\as=1}^L  \u_\as \bigg\|^2 \\
&\overset{(*)}{\leq} \int_K \bigg( \sum_{\as=1}^L  \| \u_\as(x) \| \bigg)^2
= \int_K \bigg( \sum_{\as=1}^L  \sqrt{ \nabla^2(x) \ell|_x(\u_\as,\u_\as) } \bigg)^2 \:,
\end{align*}
where in~$(*)$ we used the triangle inequality for the seminorm~\eqref{seminorm}.
As a consequence, the inequality~\eqref{var2c} is strict, which implies that also
the inequality~\eqref{var2b} is strict. This concludes the proof.
\QED

\section{Application: Hilbert Spaces of Jets} \label{sechilbert}
As an application, we now explain how our positive functionals can be used to endow the
space of jets in space-time with Hilbert space structures.
These Hilbert space structures should be very useful 
because they make functional analytic tools applicable to the 
analysis of the jet spaces and the causal action principle.
We now introduce the following bilinear forms on~$\J^\infty_0$:
\begin{align}
\lla \u, \v \rra &:= 
\int_M d\rho(x) \int_M d\rho(y) \:\nabla_{1,\u} \nabla_{2,\v} \L(x,y) 
+ \int_M \nabla^2 \ell|_x(\u,\v)\: d\rho(x) \label{sp1} \\
\llb \u, \v \rrb &:= \la \u, \v \ra + \int_M \nabla^2 \ell|_x(\u,\v)\: d\rho(x) \:. \label{sp2}
\end{align}
By Theorem~\ref{thmstable}, both bilinear forms are positive semi-definite.
Thus dividing out the null space and forming the completion gives real
Hilbert spaces of jets denoted by~$\H^{\la.,.\ra}$ and~$\H^{\la\!\la.,.\ra\!\ra}$, respectively.
Obviously,
\[ \lla \u, \u \rra \leq \llb \u, \u \rrb \:, \]
giving rise to a norm-decreasing mapping~$\H^{\la\!\la.,.\ra\!\ra} \rightarrow \H^{\la .,. \ra}$.

For the scalar components of the jets, the two scalar products~\eqref{sp1} and~\eqref{sp2} obviously agree.
But they are quite different for the vector components.
In order to understand this difference, it is instructive to consider
a jet~$\u=(0,u)$ which describes a {\em{symmetry}}
of the Lagrangian, i.e.\ (for details see~\cite[Section~3.1]{noether})
\[ \big( D_{1,u} + D_{2,u} \big) \L(x,y) = 0 \qquad \text{for all~$x,y \in M$}\:. \]
For this jet, a direct computation shows that
\[ \lla \u, \u \rra = 0 \:. \]
Hence symmetry transformations lie in the kernel of the bilinear form~$\lla ., . \rra$
and thus correspond to the zero vector in~$\H^{\la.,.\ra}$.
For example, in the setting of causal fermion systems,
jets describing global phase transformations (see~\cite[Section~5.1]{noether})
are not contained in~$\H^{\la.,.\ra}$. 
Generally speaking, the scalar product~$\lla.,.\rra$ makes it possible to
disregard symmetry transformations of the causal fermion system.
However, jets describing symmetry transformations do in general correspond to
non-zero vectors of the Hilbert space~$\H^{\la\!\la.,.\ra\!\ra}$.

\section{Application: A Positive Surface Layer Integral} \label{secosi}
In the setting of causal variational principles, the usual integrals over
hypersurfaces in space-time are undefined.
Instead, one considers so-called {\em{surface layer integrals}}.
In general terms, a surface layer integral is a double integral of the form
\beq \label{IntrOSI}
\int_\Omega d\rho(x) \int_{M \setminus \Omega} d\rho(y) \cdots\: \L(x,y) \:,
\eeq
where~$\Omega$ is a subset of~$M$ and~$\cdots$ stands for a differential operator
acting on the Lagrangian. The structure of such surface layer integrals can be understood most easily 
in the special situation that the Lagrangian is of short range
in the sense that~$\L(x,y)$ vanishes unless~$x$ and~$y$ are close together.
In this situation, we only get a contribution to the double integral~\eqref{IntrOSI}
if both~$x$ and~$y$ are close to the boundary~$\partial \Omega$.
With this in mind, surface layer integrals can be understood as an adaptation
of surface integrals to the setting of causal variational principles
(for a more detailed explanation see~\cite[Section~2.3]{noether}).

In~\cite{noether}, it is shown that there are conservation laws expressed in terms of
surface layer integrals which generalize the well-known charge and current conservation.
In~\cite{jet}, a conserved surface layer integral was found which generalizes
the {\em{symplectic form}} to the setting of causal variational principles.
We now derive a surface layer integral which is not necessarily conserved,
but which has a definite sign.
A jet~$\v \in \J$ (not necessarily with compact support) is referred to as a {\em{solution
of the linearized field equations}} if it satisfies the equation
(for details and the motivation see~\cite{jet})
\beq \label{linfield}
\nabla_{\u} \bigg( \int_M \big( \nabla_{1, \v} + \nabla_{2, \v} \big) \L(x,y)\: d\rho(y) - \nabla_\v \:\frac{\nu}{2} \bigg) = 0
\eeq
for all~$\u \in \J^\infty$ and all~$x \in M$.

The following proposition shows that there is a positive surface layer integral.
Similar as explained at the beginning of Section~\ref{sechilbert},
this can be used to endow the jet space with a Hilbert structure.
But in contrast to the scalar products in Section~\ref{sechilbert}, where the jets
were integrated over space-time, here the scalar product is given as a surface
layer integral. This should be very useful for analyzing the dynamics of jets
in space-time.

\begin{Prp} \label{prposi}
Assume that~$\v$ is a solution of the linearized field equations~\eqref{linfield}. Then
for any compact~$\Omega \subset M$, the following surface layer integral is positive,
\[ -\int_\Omega d\rho(x) \int_{M \setminus \Omega} d\rho(y) \:\nabla_{1,\v} \nabla_{2,\v} \L(x,y) \geq 0 \:. \]
\end{Prp}
\Proof Denoting the components of~$\v$ by~$\v=(b,v)$,
we evaluate~\eqref{linfield} for~$\u=\v$ and integrate over~$\Omega$.
The resulting integrals can be rewritten as follows,
\begin{align}
0 &= \int_\Omega d\rho(x) \int_M d\rho(y) \:
\nabla_{1,\v} \big( \nabla_{1,\v} + \nabla_{2,\v} \big) \L(x,y) - \frac{\nu}{2} \int_\Omega b(x)^2\, d\rho(x) \notag \\
&= \int_\Omega \nabla^2 \ell|_x(\v,\v) \: d\rho(x) + 
\int_\Omega d\rho(x) \int_M d\rho(y) \:\nabla_{1,\v} \nabla_{2,\v} \L(x,y) \notag \\ 
&= \int_\Omega \nabla^2 \ell|_x(\v,\v) \: d\rho(x) + 
\int_\Omega d\rho(x) \int_\Omega d\rho(y) \:\nabla_{1,\v} \nabla_{2,\v} \L(x,y) \label{t1} \\ 
&\qquad + \int_\Omega d\rho(x) \int_{M \setminus \Omega} d\rho(y) \:\nabla_{1,\v} \nabla_{2,\v} \L(x,y) \:.
\end{align}
Approximating the jet~$\chi_\Omega \v$ in~\eqref{t1} by smooth jets with compact support,
one finds that the integrals in~\eqref{t1} are non-negative by Proposition~\ref{prppos2}.
This gives the result.
\QED

We finally remark that in~\cite[Section~6]{action} the surface layer integral in the last proposition
is computed in Minkowski space. \\[0.5em]
\Thanks {{\em{Acknowledgments:}}
I would like to thank Magdalena Lottner and the referees for helpful comments on the
manuscript.

\providecommand{\bysame}{\leavevmode\hbox to3em{\hrulefill}\thinspace}
\providecommand{\MR}{\relax\ifhmode\unskip\space\fi MR }
\providecommand{\MRhref}[2]{%
  \href{http://www.ams.org/mathscinet-getitem?mr=#1}{#2}
}
\providecommand{\href}[2]{#2}

\end{document}